\documentclass[conference]{IEEEtran}
\IEEEoverridecommandlockouts
\usepackage{cite}
\usepackage{amsmath,amssymb,amsfonts}
\usepackage{algorithmic}
\usepackage{graphicx}
\usepackage{textcomp}

\def\BibTeX{{\rm B\kern-.05em{\sc i\kern-.025em b}\kern-.08em
    T\kern-.1667em\lower.7ex\hbox{E}\kern-.125emX}}

\usepackage{booktabs}
\usepackage{cmll}
\usepackage{combelow}
\usepackage[usenames,dvipsnames]{xcolor}
\usepackage[inline]{enumitem}
\usepackage{hyperref}
\usepackage{fontawesome}
\usepackage{microtype}
\usepackage{multirow}
\usepackage{url}
\usepackage{tikz}
\usepackage{xspace}

\usetikzlibrary{arrows,arrows.meta,decorations.markings}

\usepackage{mymacros}
\newcommand{\mbl}{\textsf{B}\xspace}
\newcommand{\mbls}{\textsf{B-spk}\xspace}
\newcommand{\msi}{\textsf{SI}\xspace}

\newcommand{\mse}{\textsf{SE}\xspace}

\newcommand{\msen}{\textsf{SE-norm}\xspace}

\newcommand{\ffa}{f_\mathsf{a}}
\newcommand{\ffs}{f_\mathsf{s}}
\newcommand{\ffv}{f_\mathsf{v}}

\begin{document}

\title{%
    Speaker disentanglement in video-to-speech conversion
    \thanks{This work was supported in part by a grant of the Romanian Ministry of Education and Research, CNCS - UEFISCDI, project number PN-III-P1-1.1-PD-2019-0918, within PNCDI III.}
}

\author{
\IEEEauthorblockN{Dan Onea\cb{t}\u{a}}
\IEEEauthorblockA{\textit{University \textsc{Politehnica} of Bucharest} \\
Bucharest, Romania \\
dan.oneata@speed.pub.ro}
\and
\IEEEauthorblockN{Adriana Stan}
\IEEEauthorblockA{\textit{Technical University of Cluj-Napoca} \\
Cluj-Napoca, Romania \\
adriana.stan@com.utcluj.ro}
\and
\IEEEauthorblockN{Horia Cucu}
\IEEEauthorblockA{\textit{University \textsc{Politehnica} of Bucharest} \\
Bucharest, Romania \\
horia.cucu@upb.ro}
}

\maketitle

\begin{abstract}
  The task of video-to-speech aims to translate silent video of lip movement to its corresponding audio signal.
  Previous approaches to this task are generally limited to the case of a single speaker,
  but a method that accounts for multiple speakers is desirable as it allows to
  \ia leverage datasets with multiple speakers or few samples per speaker; and
  \ib control speaker identity at inference time.
  In this paper, we introduce a new video-to-speech architecture and explore ways of extending it to the multi-speaker scenario: 
  we augment the network with an additional speaker-related input, through which we feed either a discrete identity or a speaker embedding.
  Interestingly, we observe that the visual encoder of the network is capable of learning the speaker identity from the lip region of the face alone.
  To better disentangle the two inputs---linguistic content and speaker identity---we add adversarial losses that dispel the identity from the video embeddings.
  To the best of our knowledge, the proposed method is the first to provide important functionalities such as 
  \ia control of the target voice and 
  \ib speech synthesis for unseen identities
  over the state-of-the-art,
  while still maintaining the intelligibility of the spoken output.
\end{abstract}


\section{Introduction}
In this paper we are concerned with learning a mapping from silent lip video of a person talking to its corresponding audio speech signal,
while also controlling the speaker identity of the output speech. 
A compelling application of this task is to enable people who lost their ability to speak to interact with a speech synthesizer in a more personalised, fast and natural way.
A fair number of video-to-speech systems have been already proposed \cite{ephrat2017icassp,kumar2018acmm,akbari2018icassp,takashima2019,vougioukas2019interspeech},
but none of them explicitly models the speaker identity
and they can synthesise speech only in a predefined speaker's voice.
When trained on multiple speakers at once,
the results usually suffer, especially when attempting to synthesise the voices of unseen speakers~\cite{vougioukas2019interspeech}.
To counter these limitations, we propose ways of explicitly incorporating speaker information.
The benefits of this approach are \ia the ability to use larger sets of training data by allowing multiple speakers present in the dataset and \ib the possibility to control the speaker identity at inference time.

In text-to-speech synthesis (TTS) systems controlling the speaker identity has been extensively explored~\cite{gibiansky2017nips,jia2018nips}.
The most common way of conditioning the output on a speaker involves a simple augmentation of the internal representations with a speaker embedding, 
be it learnt \cite{gibiansky2017nips} or transferred from an external embedding network \cite{jia2018nips}.
However, whereas the text input to the TTS system is devoid of any speaker information,
the same cannot be said about video. 
Our initial experiments revealed that a speaker-independent network is able to maintain the correct voice for each video independent of any speaker identity conditioning.
As a result, the difficulty of our video-to-speech task lies in the fact that the two sources of information---linguistic content and speaker identity---are entangled.
Ideally, we want to be able to separately specify the content (\emph {what is being said}) from the video input and the speaker information (\emph{who is speaking}).
One simple workaround would be to apply a lip-tracking algorithm, which discards any other visual input.
However, by limiting the input to a small set of data points, other essential information is lost as well (e.g. tongue or teeth positions and movement), and it is also prone to additional modelling errors.

Therefore, we attempt to strip the entire speaker information from the video stream neural encoding,
and control the speaker through a separate input to the network.
Our approach takes inspiration from work on supervised learning of disentangled representations \cite{liu2018cvpr,hsu2019icassp}.
Our main contributions are:
\ia a new video-to-speech architecture that leverages state-of-the-art components;
\ib a network adaptation to incorporate the speaker identity;
\ic a means to improve the network's disentanglement capabilities for the content and speaker information.
Code and samples are available at \url{https://speed.pub.ro/xts/}.

\begin{figure*}
    \centering
    \begin{tikzpicture}[
            node distance=2.0cm,
            font=\footnotesize,
        ]
        \newcommand{\mylabel}[1]{{\color{black!50}\scriptsize\textsf{#1}}}
        \tikzstyle{block}=[rectangle,fill=blue!20,align=center,minimum height=0.6cm];

        \node[label={above:\mylabel{video $\vb$}}] (input 1) at (0, +1.0) {\includegraphics[height=0.75cm]{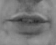}};
        \node[label={left:$t$}] (input 2) at (0,  0)   {\includegraphics[height=0.75cm]{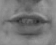}};
        \node (input 3) at (0, -1.0) {\includegraphics[height=0.75cm]{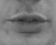}};

        \node[block, right of=input 2] (conv3d) {3D conv};
        \node[block, right of=conv3d, label={below:$\color{gray}\ffv$}] (resnet) {ResNet};
        \node[block, right of=resnet, minimum width=1.7cm] (lstm) {LSTM$_t$};
        \node[rectangle, fill=blue!40!black, right of=lstm, label=\mylabel{features}, node distance=1.75cm, minimum height=0.6cm] (feat) {};
        \node[circle, fill=blue!20, right of=feat, node distance=1.5cm, label=\mylabel{concat}] (concat) {$+$};
        \node[block, fill=orange!20, above of=feat, node distance=1.6cm, text width=1.5cm, label=$\color{gray}\ffs$] (spk clf) {speaker classifier};
        \node[right of=spk clf, label=\mylabel{speaker $\mathbf{s}$}] (output spk) {\includegraphics[height=0.7cm]{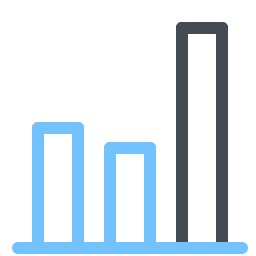}};
        \node[block, right of=concat, label={below:$\color{gray}\ffa$}] (taco) {Tacotron2\\decoder};
        \node[right of=taco, node distance=2.5cm, label={[align=center]\mylabel{audio $\ab$}}] (output) {\includegraphics[height=1.75cm]{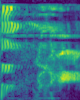}};

        \node[above of=lstm, node distance=0.8cm, minimum width=1.7cm] (lstm-prev) {\color{blue!30}LSTM$_{t-1}$};
        \node[below of=lstm, node distance=0.8cm, minimum width=1.7cm] (lstm-next) {\color{blue!30}LSTM$_{t+1}$};

        \node[block, fill=red!20, below of=lstm, node distance=1.7cm, text width=1.5cm] (spk emb) {speaker embedding};
        \node[rectangle, fill=red!40!black, right of=spk emb, minimum height=0.3cm, label={below:\mylabel{features}}, node distance=1.75cm] (feat spk) {};
        \node[left of=spk emb, label={below:\mylabel{id} \color{gray}{or} \mylabel{audio}}, align=center] (audio) {\includegraphics[height=0.6cm]{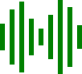}};

        \draw (input 1.east) -- (conv3d.west);
        \draw (input 2) -- (conv3d);
        \draw (input 3.east) -- (conv3d.west);

        \draw[->] (conv3d) -- (resnet);
        \draw[->] (resnet) -- (lstm);
        \draw[shorten >=2pt] (lstm) -- (feat);
        \draw[->, shorten <=2pt] (feat) -- (concat);
        \draw[->] (concat) -- (taco);
        \draw (taco) -- (output);
        \draw[->] (output.north west) ++ (0, -0.2) -| (taco);
        \draw[->, shorten <=12pt] (feat) -- (spk clf);
        \draw (spk clf) -- (output spk);

        \draw (audio) -- (spk emb);
        \draw[shorten >=2pt] (spk emb) -- (feat spk);
        \draw[->, shorten <=2pt] (feat spk) -| (concat);

        \draw[->] (lstm-prev) -- (lstm);
        \draw[->] (lstm) -- (lstm-next);
    \end{tikzpicture}
    \caption{%
        The proposed video-to-speech architecture.
        The baseline model (blue) consists of a visual processing front-end (3D conv, ResNet, LSTM) and an audio decoder (Tacotron2).
        This backbone is augmented with a speaker embedding component (red), which injects speaker information into the decoder
        (either an identity or an audio sample),
        and a speaker classifier (orange), which removes speaker information from the visual features and disentangles content and identity.
    }
    \label{fig:network}
    \vspace{-0.3cm}
\end{figure*}
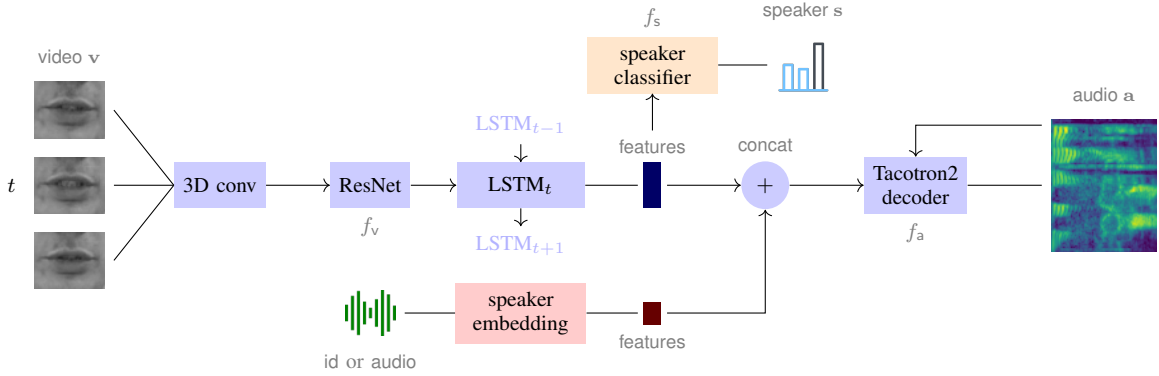

\section{Related work}
The task of video-to-speech conversion has initially been approached by using rich sensory information (ultra-sound of articulatory movements) \cite{hueber2016csl} and hand-crafted features \cite{cornu2015interspeech}.
Recent literature shifted towards deep end-to-end networks that operate directly at pixel level \cite{ephrat2017icassp,akbari2018icassp,takashima2019,vougioukas2019interspeech}.
A common aspect of all these works is the heavy use of convolutional layers,
and the notable differences rely in the output representation---%
the main choices to encode the audio are
LPC coefficients \cite{ephrat2017icassp}, spectrogram \cite{takashima2019}, or raw audio \cite{vougioukas2019interspeech}.
Recently, both Vougioukas \etal \cite{vougioukas2019interspeech} and Salik \etal \cite{salik2019aaai} evaluated the video-to-speech task in a speaker-independent scenario,
but without adapting their architectures to explicitly model the speaker identity.
The concurrent work of \cite{prajwal2020cvpr} proposes a multi-speaker architecture that incorporates the speaker embedding,
but differently from us they do not employ disentanglement techniques and
consider a scenario in which the speaker embeddings are extracted from test utterances to generalize to unseen speakers.

Two other related and well-studied tasks are lip reading and text-to-speech synthesis.
In our work we leverage state-of-the-art systems from both domains in order to build the backbone of our architecture.
More exactly,
for the visual encoder we draw inspiration from the deep lip reading front-end \cite{afouras2018interspeech}
and for the speech decoder we borrow ideas from the Tacotron2 architecture \cite{shen2018icassp}.
Both of these systems use words as either output or input, while our work avoids the explicit textual representation,
resulting in a simpler, attention-free architecture (since video and audio are time-aligned).


\section{Method description}

This section describes our methodology for the video-to-speech task.
We first describe a baseline model which we then augment with an auxiliary branch to incorporate speaker information,
and an extra loss function to encourage the disentanglement of the speaker identity from the visual features.

\textbf{Baseline model.}
The base model (\mbl) combines the ResNet architecture \cite{he2016cvpr} (for visual processing) with the Tacotron2 decoder \cite{shen2018icassp} (for speech synthesis).
As the visual and audio signals are aligned,
we do not use attention modules,
but simply upscale temporarily the visual features to match the number of audio frames.
To make the model more suitable for video processing,
we insert a layer of 3D convolutions and add an LSTM layer over the ResNet features. 
The architecture is illustrated in Figure~\ref{fig:network}.
The network generates reduced Mel spectrograms
which are inverted to audio using the Spectrogram Super-resolution Network (SSRN) \cite{Dabs-1710-08969} combined with a Griffin-Lim \cite{Griffin-Lim} vocoder.


\textbf{Modeling speaker identity.}
In order to control the voice of the synthesized audio,
we concatenate the speaker identity to the visual features (obtained after the LSTM) and
the augmented vector is then passed to the audio decoder.
The speaker information is either:
\ia a learned speaker-specific vector based on the discrete identity of each speaker (\msi) or
\ib an utterance-level embedding obtained from a pre-trained speaker recognition network (\mse).
For the \mse method, we extract embeddings using the state-of-the-art network of Xie \etal \cite{xie2019icassp}.
We freeze the speaker recognition network and project the speaker embeddings from 512 dimensions to 32 using a learned linear transformation.
The dimensionality reduction step ensures that the \mse and \msi architectures have a comparable number of parameters.
We found that the scale of the speaker embeddings affects the behaviour of the network.
In particular, for \mse, the original scale of the embeddings is small
and the model tends to ignore the auxiliary speaker input,
relying only on the information available in the visual input.
Hence, we experiment with a variant (\msen), which standardizes the speaker embedding (normalize to zero mean and unit variance).


\textbf{Disentangling identity from content.}
We hypothesize that we can control the speaker identity better if the intermediate embedding (fed to the audio decoder)
contains the content and speaker information disentangled:
ideally, the visual features should reflect only the content of what is being said,
while the auxiliary speaker embedding should contain only speaker information.
However, as mentioned in the introduction, we observed that the visual front-end consistently extracts speaker information.
We change the architecture to \textit{dispel} the speaker information from the visual features.
The idea is to train a speaker classifier on top of the features extracted from the visual processing network,
and use the classifier's predictions to update the extraction process.
We consider two approaches.
The first approach is based on adversarial learning:
a discriminator learns to classify speakers based on visual features,
while the generator changes the visual features to fool the discriminator and still be able to reconstruct the original audio.
Concretely, we optimize the following two losses:
\begin{align}
    L_\mathrm{d}(\ffs) &= H(\mathbf{s}, (\ffs \circ \ffv)(\vb))) \label{eq1} \\
    L_\mathrm{g}(\ffa, \ffv) &= \left\| \ab - (\ffa \circ \ffv)(\vb) \right\|_2^2 - \lambda H\left((\ffs \circ \ffv)(\vb)\right) \label{eq2}
\end{align}
where $\vb$ denotes the input video, $\ab$ the target audio, and $\mathbf{s}$ the speaker identity;
$f$ are the neural network functions, with the subscripts denoting the different components:
$\ffv$ the video processing net, $\ffa$ the audio decoder net, and $\ffs$ the speaker classifier.
$H$ denotes the cross-entropy (Eq. \ref{eq1}) or entropy (Eq. \ref{eq2}) and $\lambda$ represents the weight of the speaker classifier loss.
Similar losses have been proposed for face generation \cite{liu2018cvpr},
but our generator maximizes the entropy of the predictions
instead of minimizing the cross-entropy with respect to the true speaker identity.

The second approach to disentanglement uses the idea of gradient reversal \cite{ganin2015icml}.
We optimize for an objective function involving audio reconstruction and speaker classification:
\begin{align}
  L(\ffa, \ffs, \ffv) &= \left\| \ab - (\ffa \circ \ffv)(\vb) \right\|_2^2 \notag \\
                      &+ \lambda H(\mathbf{s}, (\ffs \circ \ffv)(\vb)))
\end{align}
where the notations are the same as in Eqs.~\ref{eq1} and~\ref{eq2}.
The gradient reversal step updates the visual features such that they become indistinguishable in terms of speaker identity.

Both approaches rely on a speaker classifier $\ffs$ to predict the speaker identity for the entire input video sequence.
We used two speaker classification variants, which differ in terms of complexity.
The simpler variant (denoted as \textit{linear}) performs average pooling across the temporal sequence of visual features and then applies a linear classifier.
The more complex variant (denoted as \textit{MLP}) uses an additional two-layer network before average pooling.

\section{Experimental results}

\textbf{Dataset.}
We carried our experiments on the GRID corpus~\cite{cooke2006jasa}, the test bed for the video-to-speech task \cite{ephrat2017icassp,akbari2018icassp,vougioukas2019interspeech}.
The dataset consists of 34,000 video-audio samples coming from 34 different speakers.
The vocabulary is constrained to 52 words, but no two samples contain the exact same sequence of words.

\textbf{Evaluation metrics.}
We use multiple metrics to asses the desired properties of video-to-speech systems: \emph{quality}, \emph{intelligibility}, and \emph{speaker identity}.
To measure the \emph{quality} of the generated signal, we employ 
Mel-cepstral distortion (MCD) 
and perceptual evaluation of speech quality (PESQ). 
For \emph{intelligibility},
we report the short-term objective intelligibility (STOI) 
and word error rate (WER) between the reference transcription and the output of an automatic speech recognition (ASR) system on the synthesised audio.
The ASR is implemented as a time-delay neural network \cite{peddinti2015interspeech} in Kaldi and it was trained on the TED-LIUM 2 dataset \cite{rousseau2014lrec};
the language model uses a finite state grammar derived from the word sequences present in the GRID dataset.
The performance of the ASR system is around 2.8\% WER on the natural speech samples.
Finally, to measure the \emph{identity} of the speaker we compute the equal error rate (EER) on pairs of audio---natural and synthesized.

\textbf{Implementation details.}
To improve the conditioning on the input and make the model less dependent on the auto-regressive signal,
we experimented with a dropout-like mechanism inspired by \cite{liu2019arxiv}.
The idea is to randomly replace a fraction of the audio frames on which we condition (20\% in our case) with a fixed frame---the mean of all the frames over the training set.
The weight $\lambda$ for the disentanglement losses was set to $10^{-4}$.
Other hyper-parameters (learning rate, parameters for the 3D convolution and LSTM layers) were chosen by a random search procedure on the validation split.
The SSRN vocoder was trained on 17 hours of speech data collected from multiple speakers, but not including the GRID speakers.

\subsection{Evaluation of video-to-speech synthesis}
\label{subsec:state-of-the-art}

We evaluate our proposed methods for the task of video-to-speech and compare to previously proposed methods \cite{akbari2018icassp,vougioukas2019interspeech}.
We consider the speaker dependent setup of \cite{vougioukas2019interspeech}, which consists of four speakers,
each with 900 samples for training, 50 for validation and 50 for testing.
We evaluate three variants of our methods:
a speaker-independent baseline trained on all four speakers at once (\mbl),
a speaker-dependent baseline trained for each speaker separately (\mbls),
and a model trained on all four speakers at once, but which explicitly incorporates the speaker identity (\msi).
The quantitative results are presented in Table~\ref{table:base-results},
while qualitative samples can be found online.\footnote{\url{https://speed.pub.ro/xts/}}

Compared to previous work, our methods obtain competitive results with respect to both speech quality and content intelligibility metrics,
yielding best results in terms of PESQ and WER.
The performance of our proposed methods is similar, but they come with a different set of trade-offs.
The \mbls and \msi variants model the speaker explicitly, allowing for the additional functionality of controlling the speaker identity of the generated voice.

The \msi method has the additional benefit of being more efficient than \mbls, which scales linearly in the number of speakers
as it trains a separate model for each speaker in the dataset.

Surprisingly, even if the \mbl model is completely oblivious to the explicit speaker identity,
it can still produce results on par with the speaker-dependent variants (\mbls and \msi).
Indeed, the left image of Figure~\ref{fig:embeddings} shows that network is able to implicitly model the speakers,
due to the visual cues.

\begin{table}
    \center
    \caption{
      Evaluation on seen speakers from the GRID corpus for three of our methods: baseline trained on all speakers (\mbl), baseline trained per speaker (\mbls), speaker independent model (\msi).
      The results marked with $\dagger$ denote that they were recomputed.
      Arrows indicate the direction of better performance for each metric.
    }
    \small
    \setlength{\tabcolsep}{3pt}
    \begin{tabular}{lrrrr}
        \toprule
        & STOI $\uparrow$ & PESQ $\uparrow$ & MCD $\downarrow$ & WER $\downarrow$ \\
        \midrule
        Lip2AudSpec \cite{akbari2018icassp}      & 0.446 & 1.82 & 38.14 & 32.5 \\
        V2S GAN \cite{vougioukas2019interspeech} & 0.518 & 1.71 & 22.29 & 26.6 \\
        V2S GAN \cite{vougioukas2019interspeech}$^\dagger$ & \textbf{0.525} & 1.72 & \textbf{22.02} & 27.1 \\
        \midrule
        \mbl                            & 0.470 & \textbf{1.88} &32.28 & 21.8 \\
        \mbls                           & 0.452 & 1.82 & 32.42 & \textbf{17.8} \\
        \msi                            & 0.468 & 1.85 & 32.08 & 19.9 \\
        \bottomrule
    \end{tabular}
    \label{table:base-results}
\end{table}

\subsection{Controlling the speaker identity}
\label{subsec:speaker-control}

\begin{figure}

    \centering{
    \begin{tabular}{cc}
    
        \includegraphics[width=0.15\textwidth]{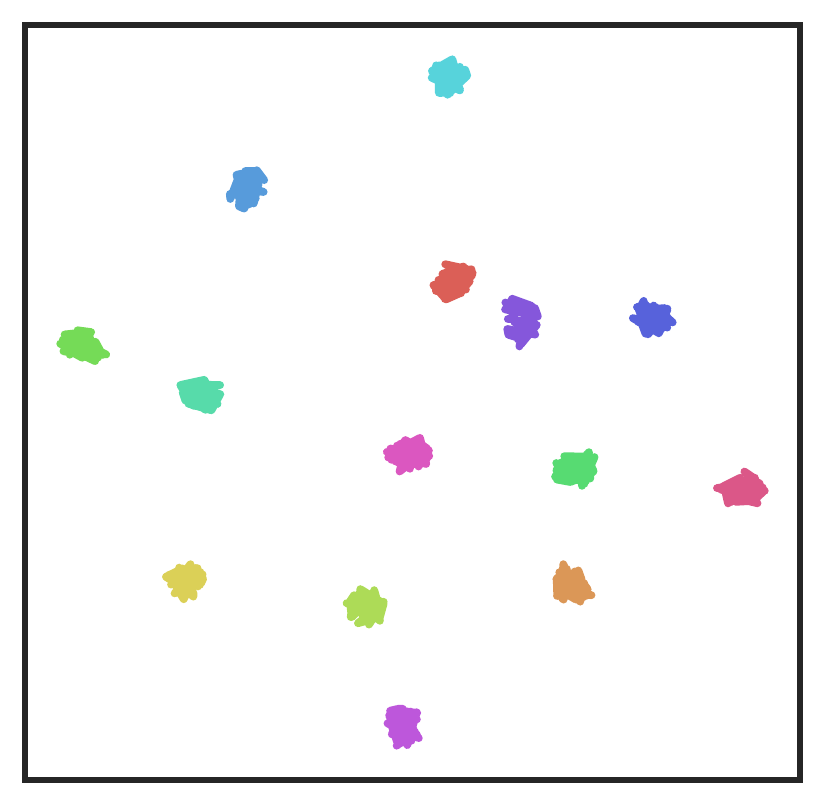} &
        \includegraphics[width=0.15\textwidth]{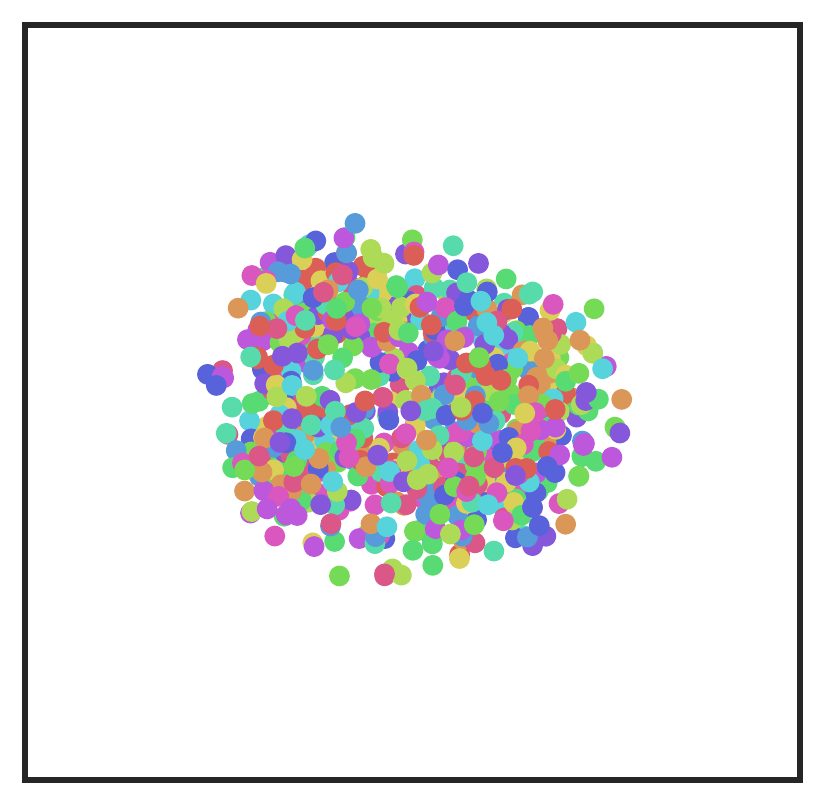} \\
        \footnotesize{speaker independent} &
        \footnotesize{speaker dependent} \\
        &
        \footnotesize{fixed speaker embedding} \\
    \end{tabular}}
    \caption{%
        Speaker embeddings for synthesised audio generated from videos of 14 different speakers (denoted by a different color).
        On the left is a speaker-independent model (\mbl) trained on 14 speakers, which 
        maintains the voice of the person in the video without using the explicit speaker identity.
        On the right is a speaker-dependent model (\msi), which controls the identity through an input speaker embedding---%
        in the figure all samples are generated in the same voice using a fixed speaker embedding (the mean of all training embeddings).
    }
    \label{fig:embeddings}
\end{figure}

In this subsection we present experiments on the control of the speaker identity for the generated audio.
The input consists of a video and a target identity,
and the desired output is an utterance consisting of the words pronounced in the video, but in the voice of the selected identity. As a qualitative example, we show in Figure~\ref{fig:embeddings} (right) the speaker embeddings extracted from the audio generated by running the \msi method with a fixed embedding (the average speaker embedding across all speakers).
As expected, even if the videos belong to different persons, the synthesised audio roughly pertains to the same speaker, which means that the average embedding can indeed control the voice identity.

For the quantitative evaluation, we consider a setup where we pair videos of \textit{unseen} speakers with target voice identities of \textit{seen} speakers.
This scenario measures the controllability of the output speech in terms of speaker identity, as well as the network's capability to encode the linguistic content from the video.
Because of the random pairing of video and speaker identities, MCD, STOI or PESQ metrics cannot be computed as all three rely on a reference audio signal.
Instead, we focus on measuring the content intelligibility and speaker identity.
The intelligibility is computed as the WER of the previously described ASR system over the output utterance.
The speaker identity of the synthesized speech is evaluated in terms of equal error rate (EER).
The EER operating point relies on the false acceptance and rejection rates given by varying a threshold on the cosine distance between the embeddings of the synthesised and natural samples;
the embeddings are obtained from the same speaker embedding network used in the training process.
This evaluation protocol measures that the embedding of the synthesized speech for a given speaker is
\ia close to all embeddings of natural samples for that speaker, and
\ib far from embeddings of any other speaker.
We randomly chose 20k pairs of embeddings out of which 1460 are positive (synthesized and natural embeddings pertain to the same speaker), and the rest are negative (the embeddings are from different speakers).
A na\"ive method, which assigns a random distance between 0 and 1 to any pair, would obtain an EER of around 50\%.

The results of this evaluation scenario are presented in Table~\ref{tab:speaker-control}.
Previous work (row 1) and our baseline (\mbl; row 2) generate significantly less intelligible speech for unseen speakers (WER 41.9\%)
than for seen speakers (Table~\ref{table:base-results}: WER 22--27\%).
Moreover, these two methods are not able to control the speaker identity,
while the rest of the architectures presented in Table~\ref{tab:speaker-control} can (rows 3–20).
In terms of speaker control, \msi yields the best results on average, 
but this architecture is only able to produce speech in the voices seen during training.
The other architectures (\mse and \msen) are capable of producing speech based on a speaker \emph{embedding}.
Among these two, the variant which uses normalized speaker embeddings (\msen) offers better speaker control,
because it ensures that the speaker embeddings have a comparable scale to the visual features.

\begin{table}
  \centering
   \addtolength{\tabcolsep}{-1.5pt}
   \renewcommand{\arraystretch}{0.9}
  \caption{%
    Word error rate (WER) and equal error rate (EER) results for the speaker identity control on unseen speakers. Arrows indicate the direction of better performance for each metric.
  }
  \small
  \newcommand{\ii}[1]{\scriptsize{\color{gray} #1}}
  \begin{tabular}{rlcccrr}
    \toprule
    & Architecture & Drop & \multicolumn{2}{c}{Disentanglement} & WER $\downarrow$ & EER $\downarrow$ \\
    \midrule

    \ii{1}   & V2S GAN \cite{vougioukas2019interspeech} & --  & –          & –      & 41.9     & N/A \\
    \ii{2}   & \mbl                                     & no  & –          & –      & 41.9     & N/A \\
    \midrule
    \ii{3}   & \multirow{6}{*}{\msi}                    & no  & –          & –      &     43.7 &     6.9 \\
    \ii{4}   &                                          & yes & –          & –      &     43.8 &     7.1 \\
    \ii{5}   &                                          & yes & dispel     & MLP    &     50.2 &     7.5 \\
    \ii{6}   &                                          & yes & dispel     & linear &     43.7 & \bf 6.8 \\
    \ii{7}   &                                          & yes & rev. grad. & MLP    &     45.2 &     6.9 \\
    \ii{8}   &                                          & yes & rev. grad. & linear & \bf 42.7 &     7.3 \\
    \midrule                                                                                
    \ii{9}   & \multirow{6}{*}{\mse}                    & no  & –          & –      &     36.5 &    18.0 \\
    \ii{10}  &                                          & yes & –          & –      & \bf 31.2 &    48.6 \\
    \ii{11}  &                                          & yes & dispel     & MLP    &     41.9 & \bf 7.1 \\
    \ii{12}  &                                          & yes & dispel     & linear &     35.5 &    12.7 \\
    \ii{13}  &                                          & yes & rev. grad. & MLP    &     37.7 &     8.9 \\
    \ii{14}  &                                          & yes & rev. grad. & linear &     36.1 &    13.6 \\
    \midrule                                                                                
    \ii{15}  & \multirow{6}{*}{\msen}                   & no  & –          & –      &     40.6 &    11.7 \\
    \ii{16}  &                                          & yes & –          & –      & \bf 38.7 &    12.5 \\
    \ii{17}  &                                          & yes & dispel     & MLP    &     49.6 &     7.8 \\
    \ii{18}  &                                          & yes & dispel     & linear &     40.1 &    10.6 \\
    \ii{19}  &                                          & yes & rev. grad. & MLP    &     41.5 & \bf 7.6 \\
    \ii{20}  &                                          & yes & rev. grad. & linear &     38.9 &    11.9 \\
    \bottomrule
  \end{tabular}
  \label{tab:speaker-control}
\end{table}

Analyzing the variants of each of the three speaker-dependent architectures,
we observe that dropping frames consistently improves intelligibility,
but this improvement is generally obtained at the expense of speaker control (rows 4, 10, 16 in Table~\ref{tab:speaker-control}). 
If we factor in the disentanglement losses (dispel and reverse gradient),
we improve the speaker control, as expected,
but lose in terms of WER.

Among the speaker classifiers used for disentanglement,
the MLP generally shows better speaker control results,
while the linear classifier yields better intelligibility results.
Overall, we noticed an on-going trade-off between the two goals (intelligibility versus speaker control),
but compared to the baseline model these results hold promise:
we are able to maintain the content intelligibility (WER 42.7\%),
while providing very good speaker control (EER 7.3\%), see row 8 in Table~\ref{tab:speaker-control}, or
we can significantly improve the results in terms of content intelligibility (WER 35.5\%),
while still obtaining decent speaker control (EER 12.7\%), see row 12 in Table~\ref{tab:speaker-control}.

\begin{figure}
    \centering{
        \includegraphics[width=\columnwidth]{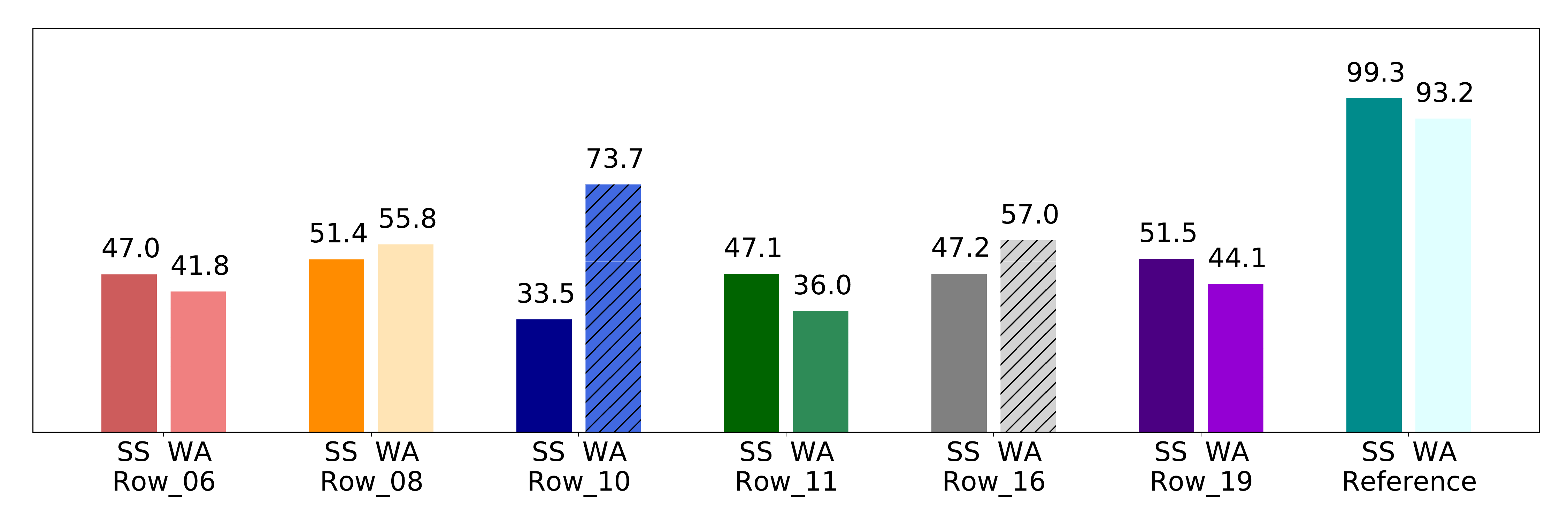} 
    }
    \vspace{-1.5em}
    \caption{
        Listening test results for six of our methods (identified by the row number in Table~\ref{tab:speaker-control}) and reference samples.
        We report speaker similarity (SS; evaluated on a MuSHRA scale, 0--100)
        and intelligibility (WA; evaluated in terms of word accuracy).
        For both metrics higher values are better. 
            }
    \label{fig:listening}
\end{figure}

\textbf{Listening test.}
To obtain subjective measures of speaker similarity (corresponding to the objective EER) and intelligibility (corresponding to the objective WER) we conducted a listening test; 
the test was carried out on eighteen volunteers and using samples from six methods from Table~\ref{tab:speaker-control} (which are identified by their row number).
The results are presented in Figure~\ref{fig:listening} and show that the relative ordering of the methods is similar for both the subjective and objective measures.
In particular, we observe that
the best performing system with respect to each of the two objective measures is best evaluated in the listening test, as well.

\section{Conclusions}

This paper addressed the task of multi-speaker video-to-speech conversion.
We highlighted the challenge of entangled content and speaker identity in the video features
and proposed ways to mitigate this issue and, consequently, allow for better speaker control.
Our methodology extended a baseline video-to-speech architecture with speaker information
by inputting speaker identities or speaker embeddings.
All these variants maintain or improve the content intelligibility,
while allowing the control of the speaker identity.
From a practical point of view, this is a highly-desirable functionality,
as an impaired user can choose a voice---either an existing voice from the database or even discover new voices by interpolating in the embedding space.
We have conducted a realistic and difficult evaluation in which we assumed we encounter new (unseen) speakers at test time.
Our ablation study showed that
dropping frames consistently improved intelligibility at the expense of speaker control
and that the disentanglement losses had a reverse effect.
While we observed that intelligibility and speaker control are often at odds,
there are still good compromises to be achieved (for example, the variants based on speaker embeddings with disentanglement)
and the proposed methods fare well when compared to the speaker-independent model.

\bibliographystyle{IEEEbib}
\bibliography{main}

\end{document}